\documentclass[twoside,leqno,twocolumn]{article}

\usepackage[letterpaper]{geometry}

\usepackage{ltexpprt}
\usepackage{hyperref}
\usepackage[referable]{threeparttablex}
\usepackage{mathtools}
\usepackage{tabu}
\usepackage{multirow}
\usepackage{makecell}
\usepackage{longtable}
\usepackage{supertabular}
\usepackage{pifont}
\usepackage{xcolor}
\usepackage{colortbl}

\usepackage{placeins}
\usepackage{booktabs}
\usepackage{blindtext}
\usepackage{caption}
\usepackage{calc}
\usepackage{todonotes}
\usepackage{algorithm}
\usepackage{algorithmicx}
\usepackage[noend]{algpseudocode}

\usepackage{csvsimple}
\usepackage{siunitx}
\usepackage{booktabs}
\usepackage{adjustbox}
\usepackage{tabularx}
\usepackage{pdflscape}
\definecolor{liturgy's}{rgb}{0.86, 0.86, 0.86}
\usepackage[boldmath,autolanguage]{numprint}
\npthousandsep{\ }

\newcommand{\ie}{i.e.\xspace}
\newcommand{\etal}{et~al.}

\def\MdR{\ensuremath{\mathbb{R}}}
\DeclarePairedDelimiter\abs{\lvert}{\rvert}%
\usepackage{amsfonts}
\newtheorem{reduction}{Reduction Rule}

\usepackage{amsfonts}

\renewcommand{\abs}[1]{\left| #1\right|}

\newcommand{\realrange}[2]{\left[#1, #2\right]}

\newcommand{\unitrange}[2]{\realrange{0}{1}}

\newcommand{\NP}{\ensuremath{\mathbf{NP}}}

\newcommand{\Oh}[1]{\mathcal{O}\!\left( #1\right)}

\newcommand{\llabel}[1]{\label{\labelprefix:#1}}
\newcommand{\labelprefix}{} %

\newcommand{\discussionsize}{\small}

\newenvironment{code}{\noindent%
\begin{tabbing}%
\hspace{2em}\=\hspace{2em}\=\hspace{2em}\=\hspace{2em}\=\hspace{2em}\=%
\hspace{2em}\=\hspace{2em}\=\hspace{2em}\=\hspace{2em}\=\hspace{2em}\=%
\kill}{\end{tabbing}}

\newcommand{\labelcommand}{}
\newsavebox{\codeparam}
\newcounter{lineNumber}
\newenvironment{disscodepos}[3]{%
\renewcommand{\labelcommand}{#2}%
\renewcommand{\captiontext}{#3}%
\sbox{\codeparam}{\parbox{\textwidth}{#3}}%
\begin{figure}[#1]\begin{center}\begin{code}\setcounter{lineNumber}{1}}{%
\end{code}\end{center}\caption{\llabel{\labelcommand}\captiontext}\end{figure}}

{\end{disscodepos}}

\newdimen\endofsize\endofsize=0.5em
\def\endofbeweis{~\quad\hglue\hsize minus\hsize
                 \hbox{\vrule height \endofsize width
\endofsize}\par}

\newcommand{\Weight}{\ensuremath{w}}
\newcommand{\MaxWeight}{\ensuremath{w^*}}
\newcommand{\MCC}{\textsf{MCC}}
\newcommand{\MWC}{\textsf{MWC}}
\newcommand{\MIS}{\textsf{MIS}}
\newcommand{\MWIS}{\textsf{MWIS}}

\newcommand{\BnB}{\textsf{B\&B}} %
\newcommand{\Degree}{\textrm{deg}}
\newcommand{\MaxDegree}{\Delta}
\DeclareMathOperator*{\argmax}{arg\,max}
\newcommand{\UpperBound}{\textsl{ub}}

\newcommand{\Inc}{\textit{inc}}
\newcommand{\Exc}{\textit{exc}}

\newcommand{\ALOOP}[1]{\ALC@it\algorithmicloop\ #1%
  \begin{ALC@loop}}
\newcommand{\ENDALOOP}{\end{ALC@loop}\ALC@it\algorithmicendloop}

\algrenewcommand\algorithmicindent{1em}

\expandafter\patchcmd\csname\string\algorithmic\endcsname{\labelsep 0.5em}{\labelsep 0em}{}{}
\makeatother
\algnewcommand\AND{\textbf{and}}
\algnewcommand\OR{\textbf{or}}
\algnewcommand\Not{\textbf{not}}

\newif\ifDoubleBlind
\DoubleBlindfalse

\begin{document}

\newcommand*\samethanks[1][\value{footnote}]{\footnotemark[#1]}

\title{\Large Improved Exact and Heuristic Algorithms for Maximum Weight Clique}
\ifDoubleBlind
\else
        \author{Roman Erhardt\thanks{Heidelberg University} \and Kathrin Hanauer\thanks{University of Vienna, Faculty of Computer Science, Vienna Austria} \and Nils Kriege\samethanks\ \thanks{University of Vienna, Research Network Data Science, Vienna, Austria} \and Christian Schulz\samethanks[1] \and Darren Strash\thanks{Department of Computer Science, Hamilton College}}
\fi
\date{}

\maketitle


\begin{abstract} \small\baselineskip=9pt We propose improved exact and heuristic algorithms for solving the maximum weight clique problem, a well-known problem in graph theory with many applications. Our algorithms interleave successful techniques from related work with novel data reduction rules that use local graph structure to identify and remove vertices and edges while retaining the optimal solution. We evaluate our algorithms on a range of synthetic and real-world graphs, and find that they outperform the current state of the art on most inputs. Our data reductions always produce smaller reduced graphs than existing data reductions alone. As a result, our exact algorithm, \texttt{MWCRedu}, finds solutions orders of magnitude faster on naturally weighted, medium-sized map labeling graphs and random hyperbolic graphs. Our heuristic algorithm, \texttt{MWCPeel}, outperforms its competitors on these instances, but is slightly less effective on extremely dense~or~large~instances.\end{abstract}

\ifDoubleBlind
\clearpage
\else
\fi

\section{Introduction}
Finding cliques in graphs is a classic problem in graph theory with many applications. In social networks, group behavior can be predicted with the help of cliques~\cite{wasserman1994social}. In biochemistry, cliques can be used to study the interaction between molecules, which can inform drug discovery~\cite{muegge2001small}. Vertex-weighted graphs, and the analogous \emph{maximum weight clique problem} (\MWC{}), can be used in an even wider variety of applications including video object co-segmentation~\cite{zhang2014video}, coding theory~\cite{zhian2013increasing}, combinatorial auctions~\cite{wu2015solving}, and genomics~\cite{butenko2006clique}.

Solving the maximum (unweighted) clique problem has been the subject of extensive research~\cite{carraghan1990exact,li2010efficient,tomita2003efficient,tomita2010simple,wu2015review,ostergard2002}, with the most effective solvers combining branch-and-bound with MaxSAT reasoning for pruning~\cite{li2017,sansegundo2022}. However, state-of-the-art algorithms still struggle to find solutions for certain instances in a reasonable time limit. Indeed, there are still unsolved instances, and recently closed instances have required over a year of computation~\cite{xiang2013scalable}. Recent work has focused on solving weighted variants of \NP{}-hard graph problems~\cite{cai2016fast,lamm2019exactly,wang2020reduction}, which are more difficult in practice.

One powerful technique for tackling \NP{}-hard graph problems is to use \emph{data reduction rules}, which remove or contract local graph structures, to reduce the input instance to an equivalent, smaller instance. Originally developed as a tool for parameterized algorithms~\cite{cygan2015parameterized}, data reduction rules have been effective in practice for computing an (unweighted) maximum independent set~\cite{chang2017computing,lamm2017finding,strash2016power} / minimum vertex cover~\cite{akiba-tcs-2016}, maximum clique~\cite{chang2020,verma2015}, and maximum $k$-plex~\cite{conte2021meta,jiang2021new}, as well as solving graph coloring~\cite{lin2017reduction,verma2015} and clique cover problems~\cite{gramm2009data,strash2022effective}, among others~\cite{Abu-Khzam2022}. However, recent work has only scratched the surface for \emph{weighted} problems. Lamm~\etal~\cite{lamm2019exactly}, Gellner~\etal~\cite{gellner2021boosting}, and Gu~\etal~\cite{gu2021towards} recently introduce an extensive collection of effective data reductions for maximum weight independent set problem (\MWIS{}), and Wang~\etal~\cite{wang2020reduction} perform data reduction for weighted graph coloring. 

However, to our knowledge, the only data reduction rules for \MWC{} remove vertices simply based on the weight of a neighborhood or the largest weight of a neighbor~\cite{cai2016fast}. Thus, there is untapped potential for reducing input instances further, making them more amenable to exact solving. One strategy is to apply \MWIS{} reductions to the \emph{complement} of the input; however, \MWIS{} reductions are most effective on large, sparse instances and the complements of the graphs considered here are dense and unlikely to fit in memory.

\textbf{Our Results.}
We develop a suite of novel exact and heuristic data reduction rules for \MWC{}, with the goal of reducing the number of vertices and edges in the input graph while maintaining solution quality. To the best of our knowledge our data reduction rules are the first to exploit local graph structures for the \MWC{} problem. We also present data reduction rules that are solely aimed at removing \emph{edges} in a graph, which to the best of our knowledge has not been done before for similar problems.
After reducing the graph, we apply either heuristic or exact algorithms on the remaining instance to obtain a solution to the original input. We extend the recent reduce-and-peel framework introduced for the \MIS{} and \MWIS{} problems, engineering methods for how and when to apply the reductions and switch to the exact solver. 
Our experiments show that our algorithms outperform the state~of~the~art.

\section{Preliminaries}
\subsection{Basic Concepts.}\label{sec:basic}
We consider a simple, weighted, undirected graph $G=(V,E,\Weight)$ with
${n=|V|}$ and ${m = |E|}$, where ${V =\{1,\dots,n\}}$ is the set of vertices, $E\subseteq \{\{u,v\}\mid u,v\in V\}$ is the set of dyadic edges, and $\Weight\colon V \to \MdR_{>0}$ is a function that assigns a positive real-valued weight to each vertex.
We extend $\Weight$ to sets, such that for $V' \subseteq V$, $\Weight(V') = \sum_{v\in V'} \Weight(v)$.
The \emph{maximum weight} of $V'$ is denoted by $\MaxWeight(V') = \max_{v\in V'} \Weight(v)$.
Two vertices $u$ and $v$ are \emph{adjacent} (also \emph{neighbors}) if $\{u, v\} \in E$.
The \emph{(open) neighborhood}~$N(v)$ of a vertex $v \in V$ is defined as ${N(v) = \{u \in V \mid \{u,v\} \in E\}}$, and its \emph{closed neighborhood} is $N[v]=N(v) \cup \{v\}$.
Both definitions extend straightforwardly to the neighborhood $N(V')$ of a set of vertices ${V' \subset V}$, \ie, ${N(V') = \cup_{v \in V'} N(v)\setminus V'}$ and $N[V'] = N(V') \cup V'$.
The \emph{degree} of a vertex $\Degree(v)$ is the number of its neighbors $\Degree(v)=\abs{N(v)}$,
and $\MaxDegree := \MaxDegree(G)$ denotes the maximum degree $\max_{v \in V}\Degree(v)$.
The \emph{complement} of $G$ is defined as ${\overline{G}=(V,\overline{E})}$, where ${\overline{E}=\{\{u,v\} \mid u, v \in V \wedge u \neq v \wedge \{u,v\}\notin E\}}$ is the set of edges not present in $G$.
The \emph{density}~$\rho := \rho(G)$ of $G$ is the ratio of the number of edges present to those that could exist, $\rho(G)=\frac{2m}{n(n-1)}$.
The subgraph \emph{induced} by the subset $V'\subseteq V$ is denoted by $G[V']=(V', E')$, where $E'=\{\{v_i,v_j\}\in E\mid v_i,v_j\in V'\}$.
A set $V'\subseteq V$ is called \emph{independent} if for all pairs of vertices~$u,v \in V'$,~$\{u, v\}\not\in E$.

A \emph{clique} is a set $Q \subseteq V$ where all vertices are pairwise adjacent.
A clique in the complement graph $\overline{G}$ corresponds to an \emph{independent set} in the original graph~$G$ and vice-versa.
The \emph{maximum weight clique problem} (\MWC{}) consists in finding a clique of maximum weight. If $\Weight \equiv 1$, we obtain the \emph{maximum cardinality clique problem} (\MCC{}) (more succinctly referred to as the maximum clique problem).
The \emph{maximum independent set problem} (\MIS{}) is that of finding an independent set of maximum cardinality, whereas the \emph{maximum weight independent set problem} (\MWIS{}) asks for an independent set of maximum total weight.
The complement of an independent set is a \emph{vertex cover}, \ie a subset ${C \subseteq V}$ such that every edge $e \in E$
is incident to at least one vertex in~$C$.
The \emph{minimum vertex cover problem}, which asks for a vertex cover with minimum cardinality, is thus complementary to the maximum independent set problem.
The maximum clique problem is also dual to the maximum independent set problem and the minimum vertex cover problem %
via the complement graph $\overline{G}$.
By extension, the weighted versions of independent set and clique
are also dual to each other.

The \emph{vertex coloring problem} asks to assign a color label~$c\in\mathbb{Z}$ to each vertex such that %
no two adjacent vertices have the same label and the number of different colors is minimal.
All vertices in a clique must receive different colors.
Thus, if a graph has a vertex coloring with $k$ colors, any clique can have cardinality at most~$k$.
All these problems are \NP{}-hard.

\subsection{Related Work.}\label{ch:RelatedWork}
This paper is a summary and extension the master thesis~\cite{maErhardt}.
A lot of research has been done for both the \MCC{} and the \MWC{} problem.
As our focus in this work is on the weighted version, we only mention results for \MWC{} and largely omit solvers and results for the cardinality version unless they were extended to the weighted case.
A detailed review on approaches for \MCC{} can be found in Wu and Hao~\cite{wu2015review} as well as in Abu-Khzam~\etal~\cite{Abu-Khzam2022} in the context of data reductions.

\subsubsection{Exact Solvers.}\label{sec:relex}

Most exact solvers for the \MCC{} use a \BnB{} framework~\cite{carraghan1990exact}, which maintains a current clique $C$ and a candidate set $P=N(C)$ of vertices for extending $C$. Fast solvers prune the search space by quickly computing a tight upper bound on the clique size that can be found by including vertices from $P$ into $C$.
One successful technique to do so is to compute a greedy heuristic vertex coloring on $G[P]$ and use the number of colors as an upper bound. 
This approach was subsequently extended to \MWC{} by Kumlander~\cite{kumlander2004new} as follows:
Given a valid vertex coloring of $G[P]$ that uses $k$ colors and partitions $V$ into color classes $\mathcal{D} = D_1 \sqcup D_2 \sqcup \dots \sqcup D_{k}$, an upper bound can be computed as $\UpperBound(\mathcal{D}) = \sum_{j=1}^{k}{\MaxWeight(D_j)}$, assuming each color class contributes a vertex of maximum weight. %

Fang~\etal~\cite{fang2016exact} were the first to implement the idea of \textsf{MaxSAT} reasoning introduced by the \MCC{} solver \texttt{MaxCLQ}~\cite{li2010efficient} for \MWC{}. 
Jiang~\etal~\cite{jiang2017exact} also rely on \textsf{MaxSAT} reasoning and contributed an efficient preprocessing step that computes an initial clique $\hat{C}$ as well as a vertex branching ordering.
It furthermore computes a simple upper bound on the maximum weight clique that each vertex $v$ can be part of as $\Weight(N[v])$ and removes $v$ if $\Weight(N[v])\leq w(\hat{C})$.
\texttt{TSM-MWC}~\cite{jiang2018two} refines the approach further with a two-stage \textsf{MaxSAT} reasoning approach that applies less expensive \textsf{MaxSAT} techniques to reduce the number of branching vertices before exhaustively looking for disjoint conflicting soft clauses.
\texttt{TSM-MWC} currently achieves the best results for a wide spectrum of graph instances, most notably large sparse real-world graph instances, and is the current state-of-the-art exact solver for maximum weight clique.

\subsubsection{Heuristic Solvers.}\label{sec:relhe}
The general scheme of a local search algorithm for \MCC{} is as follows:
A clique $C$ is constructed by starting with a single vertex and repeatedly adding vertices that are adjacent to all vertices in $C$ using some evaluation function.
Again, candidate vertices are those vertices that could potentially be added to $C$.
Once no more add operations can be performed, some vertices can be removed in an attempt to construct a larger clique.

Gendrau \etal~\cite{gendreau1993solving} proposed two algorithms for \MCC{} based on this strategy:
One is a deterministic scheme which adds the vertex with the highest degree first and when no further vertex can be added, the vertex that results in the largest set of candidate vertices is removed.
The second algorithm randomly selects which vertex to add to the current solution. %
Pullan~\cite{pullan2006phased} proposed to include a swap operator in the main search procedure.
This operator looks for a vertex that is connected to all but one vertex of the current candidate clique $C$.
Furthermore, the algorithm perturbs the current candidate clique by adding a random vertex and removing all non-adjacent vertices from the clique. %

This algorithm has been extended to \MWC{} by Pullan~\cite{pullan2008approximating} by adding a vertex which is randomly chosen only among the vertices of highest weight. %
Wang \etal~\cite{wang2016two} added a prohibition rule based on configuration checking.
Cai~\cite{cai2015balance} further improved this algorithm by using a better strategy to decide which vertex from the candidate set to add next.
This strategy works by randomly sampling $k$ different candidate vertices and choosing the best vertex with respect to some benefit estimation function.
Cai and Lin~\cite{cai2016fast} combined the algorithm with data reduction rules in their solver \texttt{FastWCLq}.
The reductions they use compute upper bounds for each vertex and remove a vertex if one of the computed upper bounds is less than the weight of the current best clique.
Every time an improved solution is found by local search, the reductions are reapplied, which in turn improves the chance of local search finding the optimal solution.

\texttt{SCCWalk4l}~\cite{wang2020sccwalk} adopts the previously seen configuration checking strategies as well as data reductions.
The authors furthermore introduce a technique called walk perturbation, which adds a random vertex to the solution when the search stagnates and removes all vertices from the candidate set
that become invalid by this perturbation.
Cai \etal~\cite{cai2021semi} improved \texttt{FastWCLq} further to also apply a reduction-and-hill-climbing method based on vertex coloring.

\texttt{SCCWalk4l} and \texttt{FastWCLq} are the current state-of-the-art for heuristic \MWC{} solvers, with the former being especially dominant in small dense networks, such as graphs from the \texttt{DIMACS} and BHOSLIB challenge~\cite{wang2020sccwalk}, and the latter showing the best results in large sparse real-world networks~\cite{cai2021semi}.

\section{Data Reductions}\label{sec:exrules}
So far, only few reductions are known that can be used for the \MWC{}.
However, especially for large instances, applying exact data reductions is a very important technique to decrease the problem size.
In general, reductions allow the classification of vertices
as either
(1) part of a solution,
(2) non-solution vertices, or
(3) deferred, \ie the decision for this vertex depends on additional information about neighboring vertices that will be obtained later.
We denote by~$\mathcal{K}$ the resulting \emph{reduced graph}, where no reduction rule applies anymore.
In the following, we review existing and introduce a large set of new reductions for the \MWC{}.

\subsection{Neighborhood Weight Reduction.}
A simple but effective reduction often seen in literature~\cite{cai2016fast,cai2021semi,jiang2018two,jiang2017exact,wang2020sccwalk}
is based on the upper bound $\Weight(N[v])$ for any clique containing $v\in V$.
\begin{reduction}[\cite{cai2016fast}]\label{rule:rule1}
Let $\hat{C}$ be the highest-weight clique found so far and let $v\in V$ s.t.\ $\Weight(N[v]) \leq \Weight(\hat{C})$.
Then %
$v$ can be removed from the graph without reducing the maximum solution weight.
\end{reduction}
The rule can be applied on a vertex $v\in V$ in $\Oh{1}$ time, given that the neighborhood weight is stored and maintained throughout the reductions.

\subsection{Largest-Weight Neighbor Reduction.}
Cai~et~al.~\cite{cai2016fast} tighten the neighborhood weight reduction rule by either including or excluding the highest weight vertex $u^*$ in the neighborhood.
\begin{reduction}[\cite{cai2016fast}]\label{rule:rule2}
Let $\hat{C}$ be the highest-weight clique found so far,
let $v\in V \setminus \hat{C}$, and let $u^* = \argmax_{u \in N(v)} \Weight(u)$. %
If
$\max\{ \Weight(N[v])-\Weight(u^*), \Weight(N[v] \cap N[u^*]) \} \leq \Weight(\hat{C})$,
then %
$v$ can be removed from the graph without reducing the maximum solution weight.
\end{reduction}
For applying the rule on a vertex $v\in V$, first its highest weight neighbor $u^*$ is identified in $\Oh{\Degree(v)}$ and then the intersection of their neighborhoods is computed in $\Oh{\min\{\Degree(v),\Degree(u^*)\}}$, resulting in overall $\Oh{\Degree(v)}$ time. 
Computing the intersection of neighborhoods is a crucial operation for the application of this reduction rule as well as several others described in the following. The running time for computing $N(u) \cap N(v)$ depends on the graph representation. Assuming constant time for checking whether two vertices are adjacent, we can iterate over the smaller set and identify those that are also adjacent to the other vertex in $\Oh{\min\{\Degree(u),\Degree(v)\}}$  time. For the application to large sparse graphs we use an adjacency list and realize the operation using indicators by iterating over the neighbors of both vertices in $\Oh{\Degree(u)+\Degree(v)}$ time.

\subsection{Twin Reduction.}
\begin{figure*}[t]
\centering
\includegraphics[width=0.4\textwidth]{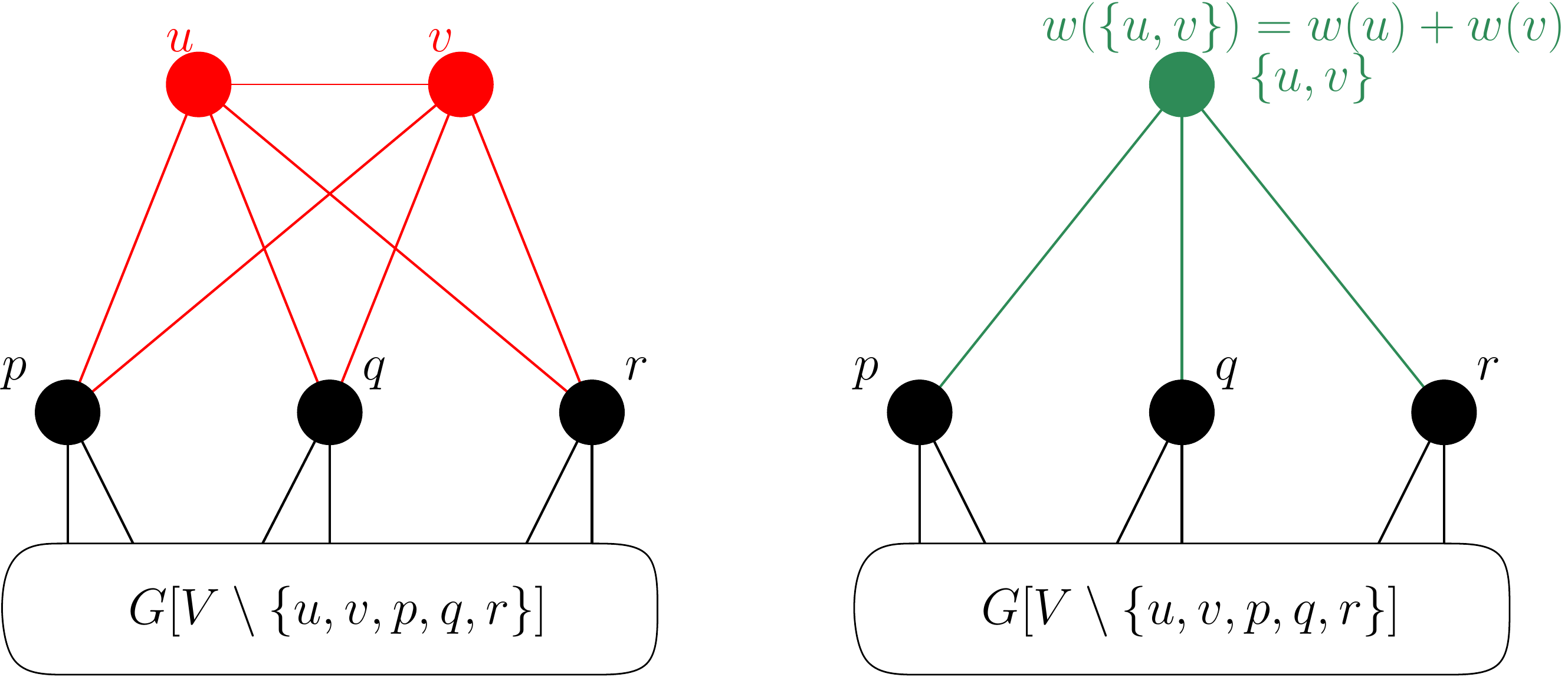} \hspace{2cm}
\includegraphics[width=0.4\textwidth]{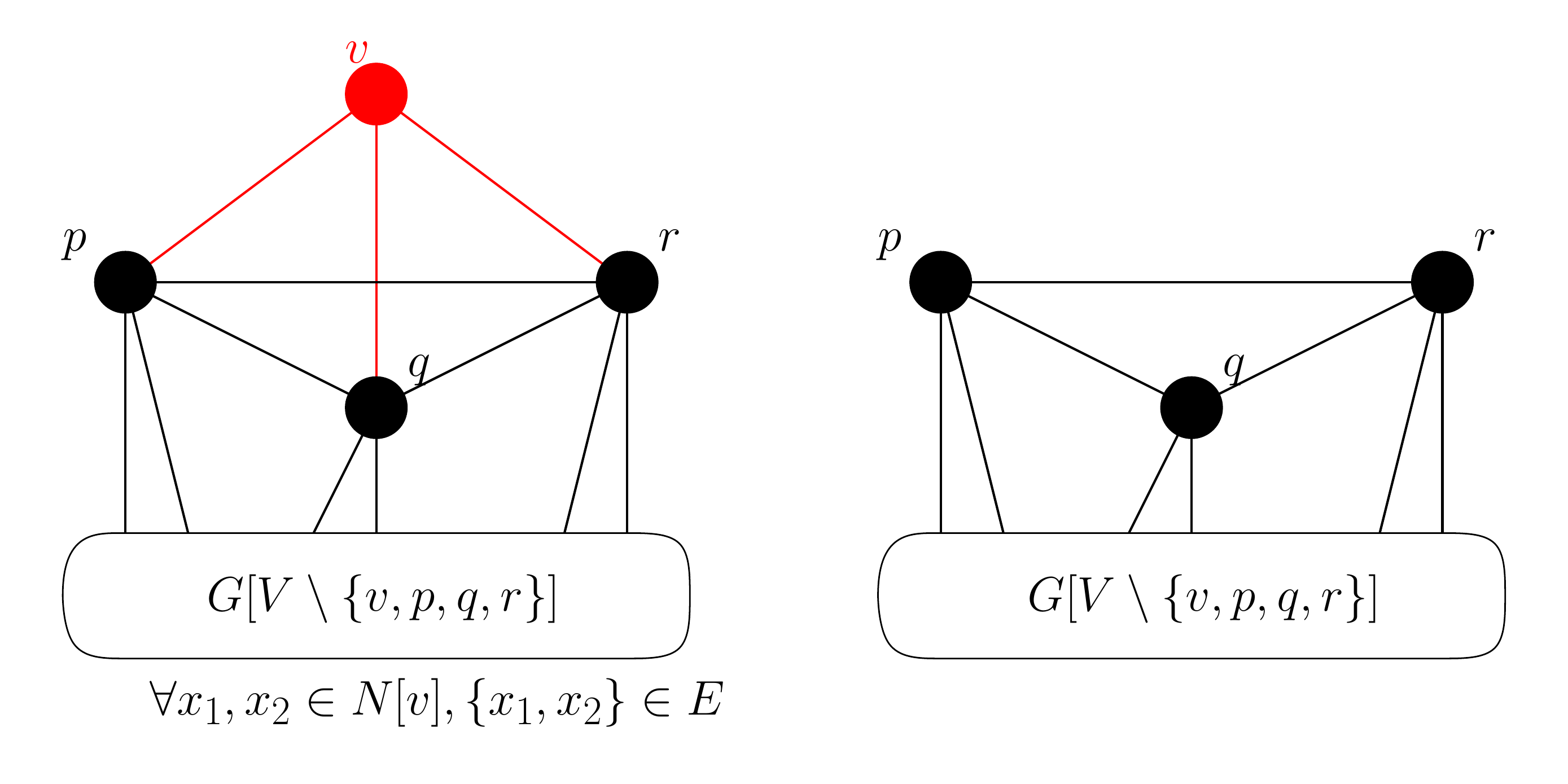}

\caption{Twin reduction (left) and simplicial vertex removal reduction (right) for \MWC{}.}%
\label{fig:mwc_twin}\label{fig:mwc_svr}
\end{figure*}

We now introduce our first new data reduction rule, based on twins.
Consider two adjacent vertices $u$ and $v$ that share the same closed neighborhood. Such vertices are called \emph{twins}.
If either one of them is in the solution, then the other one must also be in it.
Figure~\ref{fig:mwc_twin}~gives~an~illustration.
\begin{reduction}\label{rule:twin}
Let $u, v \in V$, $u \neq v$, and $N[u] = N[v]$.
Then $u$ and $v$ can be contracted to a new vertex $\{u,v\}$ with weight
$\Weight(\{u,v\}) = \Weight(u) + \Weight(v)$ and $N(\{u,v\}) = N(u) \cap N(v)$ without reducing the maximum solution weight.
\end{reduction}
\begin{proof}
Suppose there is an optimal solution $C^*$ that, w.\,l.\,o.\,g., contains $u$, but not $v$. Then it is always possible to add $v$ to the solution, as it is connected  to all neighbors of $u$, resulting in a solution of larger weight.
Hence, each optimal solution contains either both $u$ and~$v$ or neither.
\end{proof}
To check the precondition for two vertices $u, v \in V$ where $\Degree(u) = \Degree(v)$, the intersection of their neighborhoods can be obtained in time $\Oh{\Degree(v)}$ using a marking scheme.

\subsection{Domination Reduction.}
Vertex $u\in V$ is said to dominate $v\in V$ when $N(v)\subseteq N(u)$. Furthermore, if $w(v) \leq w(u)$, then
a maximal clique containing $u$ would have a weight greater or equal to one that contains $v$.
This observation leads to the following reduction rule:
\begin{figure*}[tb]
    \centering
    \includegraphics[width=0.4\textwidth]{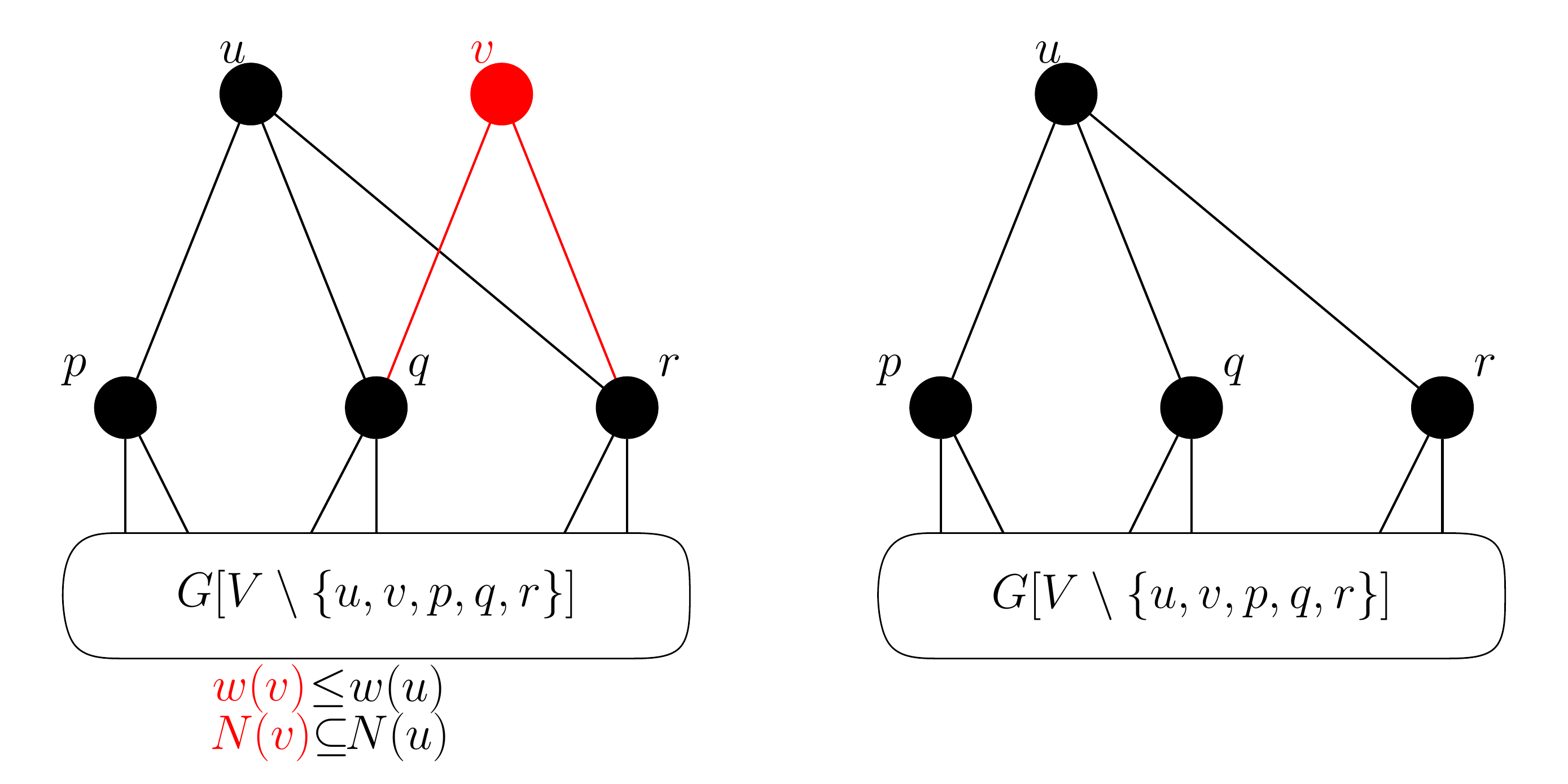}  \hspace{2cm}
    \includegraphics[width=0.4\textwidth]{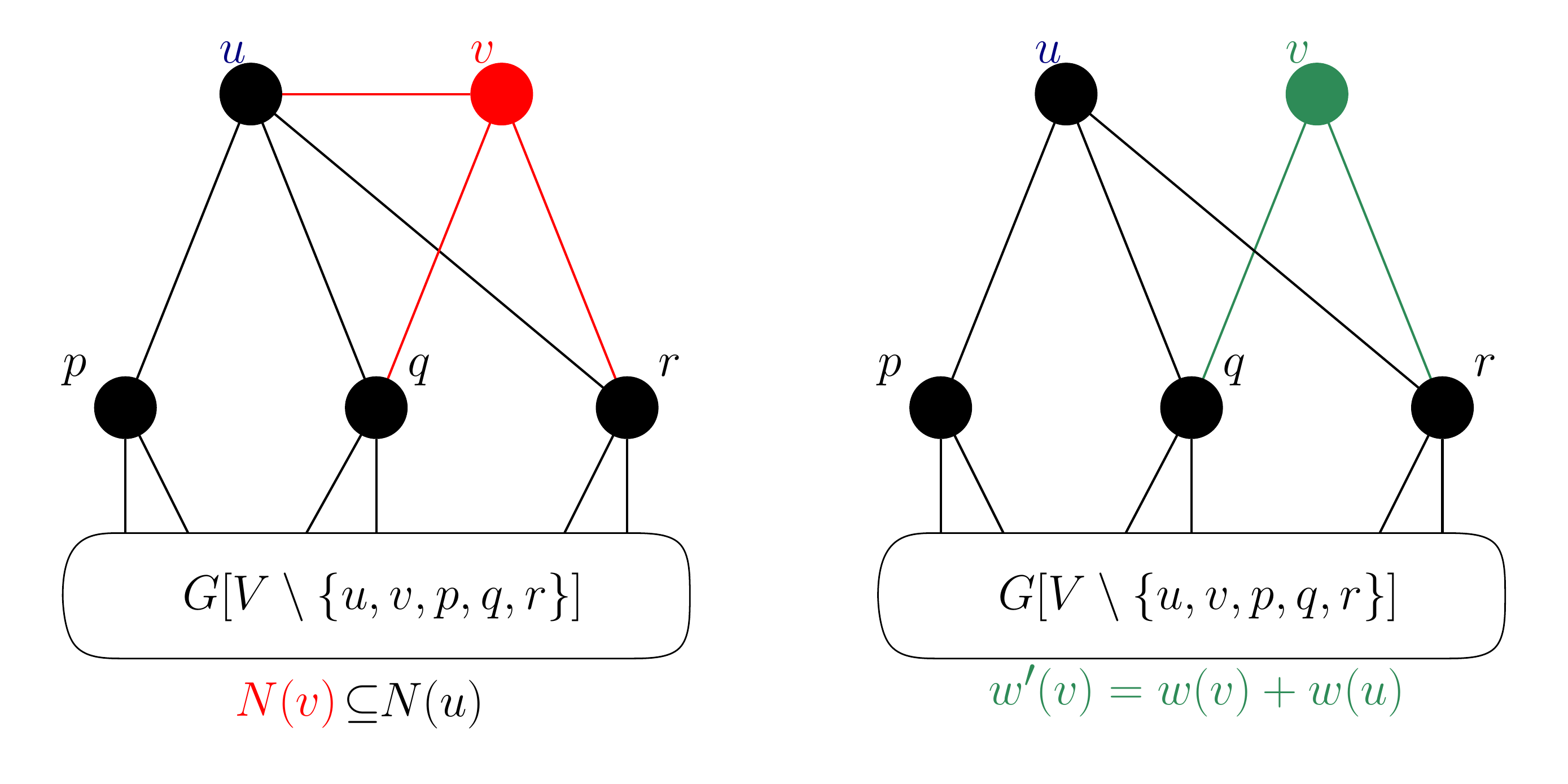}
    \caption{Domination reduction for \MWC{} by applying Reduction Rule~\ref{rule:dom1} (left) and Reduction Rule~\ref{rule:dom2} (right).}\label{fig:mwc_dom1}
\end{figure*}
\begin{reduction}\label{rule:dom1}
Let $u, v \in V$, $\{u, v\}\not\in E$, $N(v) \subseteq N(u)$, and $w(v) \leq w(u)$.
Then, $v$ can be removed from the graph without reducing the maximum solution weight.
\end{reduction}
\begin{proof}
Suppose there is an optimal solution $C^*$ that, w.\,l.\,o.\,g., contains $v$, but not $u$.
As $u$ is adjacent to all neighbors of $v$,
it is always possible to substitute $v$ with $u$ in the solution,
resulting in a solution with at least the same weight since $w(v) \leq w(u)$.
As $\{u, v\} \not\in E$, no clique can contain both $u$ and $v$.
Hence, there is at least one optimal solution that does not contain $v$.
\end{proof}
Given $v$, we find vertices $u$ with $N(u) \supseteq N(v)$ as follows:
We choose $x \in N(v)$ arbitrarily and iterate over all $u' \in N(x)$.
If $\{u', v\} \not\in E$, $\Degree(u') \geq \Degree(v)$, and $\Weight(u') \geq \Weight(v)$,
we test whether $N(v) \subseteq N(u')$ in $\Oh{\Degree(v)}$ time.
The approach identifies all vertices $u'$ for a given vertex $v$ satisfying the conditions of the reduction rule of Lemma~\ref{rule:dom1} in $\Oh{\Degree(v)\cdot\MaxDegree}$ time.

We now introduce our first reduction that is designed to remove \emph{edges} from the graph.
A similar reduction is applicable if $u$ and $v$ are adjacent.
However, simply removing $v$ is not possible, as $v$ may be part of a clique containing $u$.
Therefore, we add the weight of $u$ to $v$ and then remove the edge $\{u,v\}$, thus preserving the best solution achievable by $v$ and $u$ being in the same clique while reducing the graph at the same time.
\begin{reduction}\label{rule:dom2}
Let $u,v \in V$, $\{v,u\}\in E$, and $N(v) \subseteq N[u]$.
Then, increasing $\Weight(v)$ to $\Weight'(v) = \Weight(v) + \Weight(u)$ and removing the edge $\{u,v\}$ from the graph does not reduce the maximum solution weight.
\end{reduction}
\begin{proof}
Let $C^*$ be an optimal solution in the \emph{original} graph.
Assume that $C^*$ contains $v$, but not $u$. Then $u$ can be added to $C^*$ leading to a higher weight, contradicting the assumption that $C^*$ is optimal. Hence, if $C^*$ contains $v$, it also contains $u$.
There are two cases left to consider:\\
\textbf{Case 1:}
If $C^*$ contains both $u$ and $v$,
then $\Weight(C^*) \leq \Weight(u)+\Weight(v)+\Weight(N(v)\setminus \{u\}) = \Weight'(v) + \Weight(N(v)\setminus \{u\})$, so there exists an equivalent solution only containing $v$ in the reduced graph.
\\
\textbf{Case 2:}
If $C^*$ contains $u$ but not $v$, then $\Weight(C^*) \leq \Weight(u)+\Weight(N(u))$, and the same solution exists in the reduced graph.
\end{proof}
The reduction can be implemented analogously to the twin reduction (Reduction Rule~\ref{rule:twin}).

\subsection{Edge Bounding Reduction.}
This rule is a natural extension to  Reduction Rule~\ref{rule:rule2}, using the computed bounds not only to decide whether a vertex can be removed, but also the edge that connects it with its highest-weight neighbor.
Given a vertex $v\in V$ and its highest-weight neighbor $u^*\in N(v)$, let $\UpperBound_{\Inc}(v, u^*)$ denote the \emph{including upper bound} \mbox{$\Weight(v)+\Weight(u^*)+\Weight(N(v)\cap N(u^*))$} and let $\UpperBound_{\Exc}$ be the \emph{excluding upper bound} $\Weight(N[v]) - \Weight(u^*)$.
Reduction Rule~\ref{rule:rule2} states that $v$ can be removed if both $\UpperBound_{\Inc}(v, u^*) \leq \Weight(\hat{C})$ and $\UpperBound_{\Exc}(v, u^*) \leq \Weight(\hat{C})$,
where $\hat{C}$ is the currently best solution.
The extension provided by the edge bounding reduction is based on the observation that if $\UpperBound_{\Exc}(v, u^*) > \Weight(\hat{C})$, but $\UpperBound_{\Inc}(v, u^*) \leq \Weight(\hat{C})$, it is possible to remove the edge $\{v, u^*\}$.
We extend this rule to apply to all neighbors of $v$:
\begin{reduction}\label{rule:edge}
Let $v \in V$, $u \in N(v)$, %
and let $\hat{C}$ be the best clique found so far.
If $\UpperBound_{inc}(v, u) < \Weight(\hat{C})$, the edge $\{v, u\}$ can be removed from the graph without reducing the maximum solution weight.
\end{reduction}
\begin{proof}
The value $\UpperBound_{\Inc}(v, u)$ is an upper bound on the weight of any clique containing both $v$ and $u$.
If a clique $\hat{C}$ with weight $\Weight(\hat{C}) > \UpperBound_{\Inc}(v, u)$ is known, then there is at least one optimal solution $C^*$ that does not contain both $v$ and $u$.
The edge $\{v,u\}$ is thus irrelevant in the search for a solution of higher weight.
\end{proof}
Given an edge $\{v, u\}$, the time complexity is $\Oh{\min\{\Degree(v),\Degree(u)\}}$, as with Reduction Rule~\ref{rule:rule2}.

\subsection{Simplicial Vertex Removal Reduction.}
A vertex $v$ is called \emph{simplicial} if its closed neighborhood forms a clique $C_v$, i.e.~\mbox{$\forall x_1,x_2\in N[v]$, $\{x_1,x_2\} \in E$}.
Simplicial vertices may be removed before applying a maximum weight clique solver as well:
Once a simplex $v$ has been identified, the largest clique it can be part of is $C_v$ with $w(C_v)=w(N[v])$.
If this weight is larger than the currently known highest-weight clique, the~lower~bound~is~updated.
\begin{reduction}\label{rule:svr}
Let $v\in V$ be a simplicial vertex
and let $\hat{C}$ be the best clique found so far.
Only if \mbox{$\Weight(N[v])>\Weight(\hat{C})$}, set $\hat{C}=N[v]$.
In any case, removing $v$ from the graph then does not reduce the maximum solution weight.
\end{reduction}
\begin{proof}
If $\Weight(N[v]) \leq \Weight(\hat{C})$, $v$ cannot be part of a strictly better solution.
Otherwise, if $\Weight(N[v]) > \Weight(\hat{C})$, the same holds after the currently best solution has updated to $\hat{C}=N[v]$.
\end{proof}
Testing the adjacency of each pair of vertices in $N[v]$ takes $\Oh{\Degree(v)^2}$ in the worst case.

Observe that in contrast to the other reductions, the simplicial Vertex reduction may directly improve the currently best solution $\hat{C}$.
\begin{algorithm}[t]
\begin{algorithmic}
\Procedure{Reduce}{$G = (V, E, \Weight), \hat{C}, \textsl{lim}$}
\If{\textsf{lim}}
set vertex degree limit to $0.1\Delta$
\EndIf
\State initialize $D_i$ for each reduction rule $r_i$
\Repeat
\ForAll{reductions $r_i$}
\If{$r_i$ is not paused and $D_i \neq \emptyset$}
\State apply $r_i$ on all vertices in $D_i$
\Comment{Section~\ref{sect:applyreductions}}
\State update $D_i$
\If{reduction rate not achieved}
\State pause $r_i$
\EndIf
\State update $\hat{C}$ via local search
\Comment{Section~\ref{sec:lb}}
\ElsIf{$r_i$ paused $\wedge$ $G$ reduced enough}
\State unpause $r_i$
\Comment{Section~\ref{sect:applyreductions}}
\EndIf
\EndFor
\If{\textsl{lim} and all reductions paused}
\State increase limit on vertex degrees, update $D_i$'s
\EndIf
\Until{all reductions paused and degree unlimited}
\State\Return $G, \hat{C}$
\EndProcedure
\end{algorithmic}
\caption{Reduce graph $G$ via data reductions and improve clique $\hat{C}$,
using vertex degree limits if \textsl{lim} is set.}\label{alg:reduce}
\end{algorithm}

\subsection{Applying the Reductions.}\label{sect:applyreductions}
For applying the exact reduction rules proposed in this section, an adapted version of the strategy from Hespe \etal~\cite{hespe2019scalable} that entails both dependency checking and reduction tracking is used.
Specifically, the set of reductions $\{r_i\}$ is iterated, where each rule $r_i$ is tried on its set of viable vertices $D_i$, which is initially set to $D_i=V$. After preliminary experiments, we settled on the following order of reductions: neighborhood weight, twin, simplicial vertex, edge bounding (which includes largest-weight neighbor), domination case 1, and domination case 2.
Every time a rule $r_i$ fails to \emph{reduce} a vertex, \ie to remove it from the graph, this vertex is removed from the set of viable candidates $D_i$.
Otherwise, the set of each rule $r_j$ is updated to $D_j=D_j\cup N(v)$ and the applicable vertices or edges are removed from the graph.
This minimizes redundant computations without affecting the final size of the reduced graph~\cite{hespe2019scalable}.

\emph{Reduction tracking} aims at tracking the effectiveness of reductions.
Slightly different from the original strategy, reduction tracking is implemented by pausing a reduction once it fails to achieve a reduction rate of at least \SI{1}{\percent} of the current number of vertices or edges per second, until other reductions reduced the graph by that amount.
Reduction tracking is checked both in between the application of different reduction rules as well as periodically during the iteration over candidate vertices, in order to prevent single reductions to delay the solver and allow either more efficient reductions or the exact solver to take over.
Another addition to the strategy by Hespe~\etal{} is to set a dynamic limitation on the degree of vertices that are tried in the reductions.
The limit is set to \SI{10}{\percent} of the highest degree initially and is increased by \SI{10}{\percent} whenever the reductions have been exhaustively applied in the previous level.
This guarantees that reductions applicable on low degree vertices, which are typically more efficient, are applied first.
The loop terminates once the degree is no longer limited and all reductions are paused, at which point we run either an exact or heuristic solver on the reduced graph.

\section{\texttt{MWCRedu}: A New Exact Algorithm}\label{sec:exsolv}

Our exact algorithm \texttt{MWCRedu} works in two stages:
First, the set of exact reduction rules from Section~\ref{sec:exrules} is used to reduce the graph.
Second, the reduced graph is passed to an exact \BnB{} solver to compute the final solution.
\subsection{Computing a Lower Bound.}\label{sec:lb}
Reduction Rules~\ref{rule:rule1}, \ref{rule:rule2} and \ref{rule:edge} depend on the currently best solution $\hat{C}$ to be applicable.
For computing bounds, fast heuristics are generally preferred, since spending more time on improving the initial solution typically gives diminishing returns.
A well-suited heuristic for computing an initial lower bound is the one employed in Jiang \etal~\cite{jiang2017exact}:
Repeatedly remove the vertex with the smallest vertex degree from the graph until %
all remaining vertices are pairwise adjacent and form the initial clique $\hat{C}$,
which yields an initial lower~bound~of~$\Weight(\hat{C})$.

Afterwards, $\hat{C}$ is continuously improved by the simplicial vertex reduction (Reduction Rule~\ref{rule:svr}) and the local search algorithm from \texttt{FastWCLq}~\cite{cai2021semi}, the latter being applied on the reduced graph in between checking each reduction rule. %
Subsequently, $\hat{C}$ provides the lower bound in the Reduction Rules~\ref{rule:rule1}, \ref{rule:rule2} and \ref{rule:edge},
and it also serves as the initial solution for the solver that is applied on the reduced graph.
Algorithm~\ref{alg:reduce} gives an outline.

\subsection{Branch and Bound.}
The reduced graph is solved using the branch and bound paradigm.
As the procedure has exponential time complexity, it is important to choose a good ordering and to reduce the set of branching vertices by computing tight upper bounds.
We use the same ordering as Jiang~\etal~\cite{jiang2017exact}, \ie the ordering of the vertices is given as  $v_1<v_2<...<v_n$, where $v_1$ has the smallest vertex degree, $v_2$ has the smallest vertex degree after $v_1$ is removed, etc. Such an ordering is called a \emph{degeneracy ordering} of the graph.

To compute tight upper bounds and reduce the set of branching vertices, we apply efficient \MIS{}- and \textsf{MaxSAT}-based approaches from~\cite{jiang2017exact,jiang2018two} throughout the search.
Recall from Section~\ref{ch:RelatedWork} that for any vertex coloring that partitions $V$ into color classes $\mathcal{D} = D_1 \sqcup D_2 \sqcup \dots \sqcup D_k$,
each color class forms an independent set and $\UpperBound(\mathcal{D}) = \sum_{j=1}^{k}{\MaxWeight(D_j)}$ is an upper bound on the maximum clique weight.
The set of branching vertices is then further reduced via the two-stage \textsf{MaxSAT} reasoning approach from \texttt{TSM-MWC}~\cite{jiang2018two}.

In the first stage, which the authors refer to as binary \textsf{MaxSAT} reasoning, the set of branching vertices is reduced by inserting as many vertices as possible into the independent sets s.t.\ $\sum_{j=1}^{k'}{\MaxWeight(D_j)}\leq \Weight(\hat{C})$.
As these vertices cannot form a clique with a weight larger than $\Weight(\hat{C})$ by themselves, they can be removed from the set of branching vertices.
If a vertex $v_i\in V$ has neighbors in all existing independent sets but \mbox{$\UpperBound + \Weight(v_i)\leq \Weight(\hat{C})$} holds, it is inserted as a new independent set.
Otherwise we try to split its weight among independent sets that do not contain any of its neighbors by adding $v_i$ with weight $\MaxWeight(S_j)$ into independent $S_j$ and updating the weight to $\Weight(v_i) = \Weight(v_i) - \MaxWeight(S_j)$ for $j={1,2,...,k'}$, until its remaining weight is given as $\delta=\Weight(v_i)-\sum_{j=1}^{k'}{\MaxWeight(S_j)}$.
If $\delta > 0$ and $\UpperBound + \delta \leq \Weight(\hat{C})$, $v_i$ is inserted as a new independent set with weight $\delta$, otherwise the weight splitting procedure is undone and $v_i$ is kept in the set of branching vertices.

In the second stage, called ordered \textsf{MaxSAT} reasoning, the set of branching vertices is reduced further by detecting disjoint conflicting subsets of independent sets. Firstly, the weight of a branching vertex $v_i$ is again split among the independent sets $\{S_1,S_2,...,S_{k'}\}$ that do not contain any of its neighbors, resulting in the remaining weight $\Weight(v_i)=\delta>0$, since the vertex was not removed from the set of branching vertices in the first stage.
After that, the algorithm tries to find a set of independent sets $\{U_1,U_2,...,U_r\}$ that each contain exactly one neighbor $u$ of $v_i$.
It then looks for an independent set $D_q$ s.t.\ $D_q \cap N(v_i)\cap N(u)=\emptyset$ for any $U_j$, proving that the sets $\{\{v_i\},U_j,D_q\}$ are conflicting.
In this case, $\UpperBound$ can be further improved to $\UpperBound + \delta - \beta$, where $\beta =\min(\delta,\MaxWeight(U_j),\MaxWeight(D_q))$~\cite{jiang2018two}. 

Finally, if after considering all $U_j \in \{U_1,U_2,...,U_r\}$ $\UpperBound$ is still higher than the lower bound, $\UpperBound$ is reduced by identifying conflicting subsets via unit propagation as first implemented for maximum weight clique~\cite{fang2016exact}.
Unit propagation works from the idea that clauses with more literals are more likely to be satisfied and are thus considered \emph{weaker} clauses.
A unit clause is thus the strongest clause since it only has one possibility of evaluating to \texttt{true}.
The algorithm repeatedly satisfies such a clause, removing all occurrences of the contained literal from the other clauses.
If an empty clause remains, the set of clauses is identified as conflicting.
Each time a set of conflicting clauses $\{S_0,S_1,...,S_r\}$ is identified, the upper bound can be reduced by $\delta = \min\{\MaxWeight(S_1),\dots,\MaxWeight(S_r)\}$.
To tighten the bound further, each $S_j$ ($0\leq j \leq r$) is split into $S_j'$ and $S_j''$ so that $\MaxWeight(S_j')=\delta$ and \mbox{$\MaxWeight(S_j'')=\MaxWeight(S_j)-\delta$}.
$S_j'$ then represents the conflicting subset found so far, whereas further conflicts can be deduced from $S_j''$~\cite{fang2016exact}.

The procedure is run at every branch of the solver in order to reduce the amount of work to be done.
The algorithm terminates when all branches are either explored or pruned or when the time limit is reached, in which case the best solution found  is reported.

\section{\texttt{MWCPeel}: A New Heuristic Algorithm}\label{sec:heur}
For our new heuristic algorithm \texttt{MWCPeel}, we investigate vertex peeling techniques, which remove vertices from the graph that are assigned the lowest scores by some heuristic rule.
This rule must therefore capture the likelihood of a vertex belonging to the solution as well as possible.
Using the vertex degree is an obvious choice for \MCC{}, since a vertex with a high degree is more likely to form a large clique.
Furthermore, a vertex $v$ cannot be part of a clique larger than $\Degree(v)$.
For the measure to remain an upper bound in the context of \MWC{}, the weight of the neighborhood of each vertex is taken into account.
The resulting simple and intuitive scoring measure $\Weight(N[v])$ is used in
our peeling step.

\begin{algorithm}[t]
\begin{algorithmic}
\Procedure{MWCPeel}{$G = (V, E, \Weight)$}
\State compute initial clique $\hat{C}$
\Comment{Section~\ref{sec:lb}}
\Repeat
\Comment{Algorithm~\ref{alg:reduce}}
\State $G, \hat{C} \gets$ \Call{Reduce}{$G, \hat{C}, \textsl{isFirstIteration}$}
\State  $\mathcal{N} \gets$ \#vertices to peel off
\Comment{Section~\ref{sect:numpeeling}}
\State remove $\mathcal{N}$ vertices $v$ with lowest score $w(N[v])$
\Until{stopping criteria met}
\Comment{Section~\ref{sect:stoppeeling}}
\State \Return \Call{TSM-MWC}{$G, \hat{C}$}
\EndProcedure
\end{algorithmic}
\caption{Heuristic Solver \texttt{MWCPeel}}\label{alg:generalredualgorithm}
\end{algorithm}
Overall, our heuristic solver works similarly to the exact approach \texttt{MWCRedu} described in Section~\ref{sec:exsolv}, but implements the peeling reduction on top of the previously introduced exact reductions:
We first run exact reductions exhaustively.
On the reduced graph, we apply our peeling strategy that removes vertices that are unlikely to be part of a large clique.
We repeat the process until the remaining graph is small or the scores of the peeling reductions are not sufficiently large, and then apply the exact algorithm on the remaining~graph. Algorithm~\ref{alg:generalredualgorithm} gives an overview.%

\subsection{Peeling Strategy.}\label{sect:numpeeling}
Chang~\etal~\cite{chang2017computing} introduced a reduce-and-peel heuristic technique to repeatedly remove the minimum degree vertex from a graph, adding it to a growing independent set. For \MWC{}, a straightforward approach is to remove the vertices with the lowest score and exclude them from the solution.
More precisely, we remove a fixed percentage of the currently remaining vertices in each peeling step.
The number of vertices to be peeled off in one step, $\mathcal{N}$, is dynamically determined as follows:
\[
\mathcal{N}=
\begin{cases}
	0.1n& \text{if } n > \numprint{50000}, \\
	\max\{0.01 n,\frac{0.01}{\numprint{50000}}n\}& \text{otherwise},
\end{cases}
\]
where $n$ always refers to the current number of vertices
and the threshold of \numprint{50000} has proven itself suitable in preliminary experiments.
Without the differentiation between larger and smaller graphs, the exact reductions would often be reapplied on many vertices, which would significantly slow down the solver.
Furthermore, as the vertex degrees often follow a power-law distribution in real-world graph instances~\cite{gugelmann2012random}, the size of the optimal solution makes up a smaller portion of the graph for large graphs.
After each peeling step, the viable candidate sets are updated and exact reductions are applied again.

\subsection{Stopping Criteria.}\label{sect:stoppeeling}
Another important decision is when to stop applying the peeling reduction; stopping too early could result in a much higher running time for the solver applied on the reduced graph, whereas stopping late might negatively impact the solution quality.
Since the optimal amount of vertices to reduce is highly dependent on the graph structure, a static stopping criterion is unlikely to be a good strategy.
For this reason, we employ a dynamic strategy that works by comparing the current computed score with previously~computed~scores.

The first stopping criterion is the deterioration of the maximum score value below a certain threshold relative to the total maximum score value.
This indicates that the peeling reduction begins to reduce the maximum solution.

A second stopping criterion takes effect if the difference between the minimum and maximum score shrinks below a certain threshold.
This shows that the scoring model can no longer clearly distinguish high quality vertices from low quality vertices.

We set both thresholds to \SI{90}{\percent} to achieve a good balance between speed-up and solution quality.
As a fail-safe, a backup of the current graph state is created before applying the heuristic reduction, which can be reloaded in the case the graph is reduced to zero.
After the reduction procedure, the branch-and-bound solver is applied on the reduced graph to obtain the final result.
\section{Experimental Evaluation}

We implemented our new solvers \texttt{MWCRedu} and \texttt{MWCPeel} and evaluate them against the state-of-the-art solvers
in their class on an extensive and diverse set of instances.
More precisely, we compare our exact solver \texttt{MWCRedu} with the currently best exact solver \texttt{TSM-MWC} on each dataset,
and our heuristic solver \texttt{MWCPeel} with the currently best heuristic solvers \texttt{FastWCLq} and \texttt{SCCWalk4l}.

\paragraph*{Methodology.}
The experiments were run on an Intel Xeon Silver 4216 CPU @2.10GHz with 16 cores under Linux with 95~GB of RAM.
All solvers are implemented in C/C++ and compiled using GNU \texttt{g++} with full optimization (\texttt{-O3}).
Each solver was executed on up to 16 graph instances in parallel.
As the solvers were run exclusively on the machine, there is no relevant difference to solving the graph instances sequentially.
We always report the solution quality $\Weight(\hat{C})$ and the time to find that solution $t_{sol}$.
For exact solvers, we additionally give the time needed to prove optimality of the solution $t_{prv}$.
Solvers that use random number generation are run five times with different seeds and we report their average solutions to better capture their general performance.
If an exact algorithm did not finish within a time limit of \numprint{3600}~seconds,
it is halted and the best solution found so far is output.
Heuristic algorithms are stopped after \numprint{1000} seconds.

\paragraph*{Instances.}
We evaluate our algorithms on a broad selection of graphs, covering different sizes, densities, weightings and areas of application.
Some of the graphs are originally unweighted and thus were assigned weights artificially.
For each unweighted graph, weights are drawn uniformly from~the~range~$[1,200]$.\footnote{Other weight distributions such as power-law and exponential gave similar results and were excluded due to space constraints.}

We compiled four sets of instances,
with \numprint{58} instances altogether:
\texttt{OSM} contains \num{12} naturally-weighted map labeling instances from Cai~\etal~\cite{cai2018improving}, generated from OpenStreetMap data using the technique of Barth~\etal~\cite{barth2016temporal}.
The \num{10} instances in \texttt{REP} are real-world graphs from the network data repository~\cite{nr-aaai15},
and the \num{23} instances in \texttt{DIMACS} were taken from the second DIMACS implementation challenge~\cite{johnson1996cliques}.
Moreover, we use \num{13} random hyperbolic graphs (\texttt{RHG}).
These are randomly generated graphs such that the vertex degrees follow a power-law distribution~\cite{penschuck2020recent}
and were generated by the KaGen framework~\cite{funke2019communication}.
We varied the power-law exponent between \num{1.75} and \num{2.25} and chose the average degree between \num{100} and \num{500}.
For \texttt{REP}, \texttt{DIMACS}, and \texttt{RHG}, we assigned artificial weights as described above.
See Table~\ref{tab:properties} in Appendix~\ref{prop:instances} for detailed per-instance statistics.

\subsection{Impact of New Data Reduction Rules.}
We first investigate the impact of the reduction rules on the instances and compare the effect of adding our ``new'' rules to the ``old'' ones that are described in current literature.
Table~\ref{tab:kernelsizes} shows reduced graph sizes on all instances, and Table~\ref{tab:kernelsizessummary} shows reduced graph sizes for a subset of instances.

On the \texttt{DIMACS} instances, the new data reduction rules do not help to compute smaller reduced graphs (hence they are excluded from the table). This is expected as these instances are dense and data reduction rules tend to work well on sparse instances. On the other instances, reduced graphs are significantly smaller when the new data reduction rules are employed additionally.
\begin{table*}[t!]
	\centering
	\begin{tabular}{l@{\hskip 43pt}rrrr}
		\toprule
		& \multicolumn{4}{c}{Reduced Graph Size}  \\
    \cmidrule(lr){2-5}
		& \multicolumn{2}{c}{old+new reductions} & \multicolumn{2}{c}{old reductions only} \\
    \cmidrule(lr){2-3}\cmidrule(lr){4-5}
		Graph & absolute & \% of $n_0$ & absolute & \% of $n_0$ \\

		\midrule
		\multicolumn{5}{c}{\texttt{REP}}\\
		\midrule

		\texttt{bio-human-gene1} & \nprounddigits{0}$\textbf{\numprint{3915.46666666667}}$ & \nprounddigits{2}$\textbf{\numprint{17.5715418330865}}$ & \nprounddigits{0}$\numprint{4485.33333333333}$ & \nprounddigits{2}$\numprint{20.128947329055}$ \\ 
		\texttt{sc-TSOPF-RS-b2383} & \nprounddigits{0}$\textbf{\numprint{16122.6666666667}}$ & \nprounddigits{2}$\textbf{\numprint{42.2933990888662}}$ & \nprounddigits{0}$\numprint{37736.6666666667}$ & \nprounddigits{2}$\numprint{98.9918067906578}$ \\ 
		\texttt{soc-orkut} & \nprounddigits{0}$\textbf{\numprint{1264963.2}}$ & \nprounddigits{2}$\textbf{\numprint{42.2053099494656}}$ & \nprounddigits{0}$\numprint{1521403.66666667}$ & \nprounddigits{2}$\numprint{50.7614081658029}$ \\ 
		\texttt{web-wikipedia\_link\_it} & \nprounddigits{0}$\textbf{\numprint{0}}$ & \nprounddigits{2}$\textbf{\numprint{0}}$ & \nprounddigits{0}$\numprint{1213.66666666667}$ & \nprounddigits{2}$\numprint{0.0413315924344001}$ \\ 
		\texttt{web-wikipedia-growth} & \nprounddigits{0}$\textbf{\numprint{83723.8666666667}}$ & \nprounddigits{2}$\textbf{\numprint{4.47551286231787}}$ & \nprounddigits{0}$\numprint{637482.666666667}$ & \nprounddigits{2}$\numprint{34.0770438318428}$ \\ 

		\midrule
		\multicolumn{5}{c}{\texttt{RHG}}\\
		\midrule
		\texttt{rhg\_250k\_100\_1.75} & \nprounddigits{0}$\textbf{\numprint{7.06666666666667}}$ & \nprounddigits{2}$\textbf{\numprint{0.00282666666666667}}$ & \nprounddigits{0}$\numprint{1061}$ & \nprounddigits{2}$\numprint{0.4244}$ \\ 
		\texttt{rhg\_500k\_500\_2.25} & \nprounddigits{0}$\textbf{\numprint{0.266666666666667}}$ & \nprounddigits{2}$\textbf{\numprint{0}}$ & \nprounddigits{0}$\numprint{1761}$ & \nprounddigits{2}$\numprint{0.3522}$ \\ 
		\texttt{rhg\_750k\_250\_2.25} & \nprounddigits{0}$\textbf{\numprint{15.2}}$ & \nprounddigits{2}$\textbf{\numprint{0.00202666666666667}}$ & \nprounddigits{0}$\numprint{1061.66666666667}$ & \nprounddigits{2}$\numprint{0.141555555555556}$ \\ 
		\texttt{rhg\_750k\_500\_1.75} & \nprounddigits{0}$\textbf{\numprint{4444.73333333333}}$ & \nprounddigits{2}$\textbf{\numprint{0.592631111111111}}$ & \nprounddigits{0}$\numprint{7341.33333333333}$ & \nprounddigits{2}$\numprint{0.978844444444444}$ \\ 
		\texttt{rhg\_750k\_500\_2.25} & \nprounddigits{0}$\textbf{\numprint{11.8}}$ & \nprounddigits{2}$\textbf{\numprint{0.00157333333333333}}$ & \nprounddigits{0}$\numprint{2651.33333333333}$ & \nprounddigits{2}$\numprint{0.353511111111111}$ \\ 
		
		\midrule
		\multicolumn{5}{c}{\texttt{OSM}}\\
		\midrule
		\texttt{district-of-columbia-AM2} & \nprounddigits{0}$\textbf{\numprint{0}}$ & \nprounddigits{2}$\textbf{\numprint{0}}$ & \nprounddigits{0}$\numprint{759}$ & \nprounddigits{2}$\numprint{5.58211370155181}$ \\ 
		\texttt{greenland-AM3} & \nprounddigits{0}$\textbf{\numprint{0}}$ & \nprounddigits{2}$\textbf{\numprint{0}}$ & \nprounddigits{0}$\numprint{1768}$ & \nprounddigits{2}$\numprint{35.4592860008023}$ \\ 
		\texttt{idaho-AM3} & \nprounddigits{0}$\textbf{\numprint{0}}$ & \nprounddigits{2}$\textbf{\numprint{0}}$ & \nprounddigits{0}$\numprint{2293}$ & \nprounddigits{2}$\numprint{56.4222440944882}$ \\ 
		\texttt{massachusetts-AM3} & \nprounddigits{0}$\textbf{\numprint{0}}$ & \nprounddigits{2}$\textbf{\numprint{0}}$ & \nprounddigits{0}$\numprint{802}$ & \nprounddigits{2}$\numprint{21.658115041858}$ \\ 
		\texttt{virginia-AM3} & \nprounddigits{0}$\textbf{\numprint{0}}$ & \nprounddigits{2}$\textbf{\numprint{0}}$ & \nprounddigits{0}$\numprint{907}$ & \nprounddigits{2}$\numprint{14.6645109135004}$ \\ 
		
		\bottomrule
	\end{tabular}
        \caption{Selected instances and reduced graph sizes (number of nodes) when both old and new data reductions rules are applied vs. reduced graph sizes obtained when only running reductions from the current literature. Smaller is better. $n_0$ refers to the initial number of nodes.}\label{tab:kernelsizessummary}
\end{table*}
\begin{figure}[t]
\centering
\includegraphics[width=\linewidth]{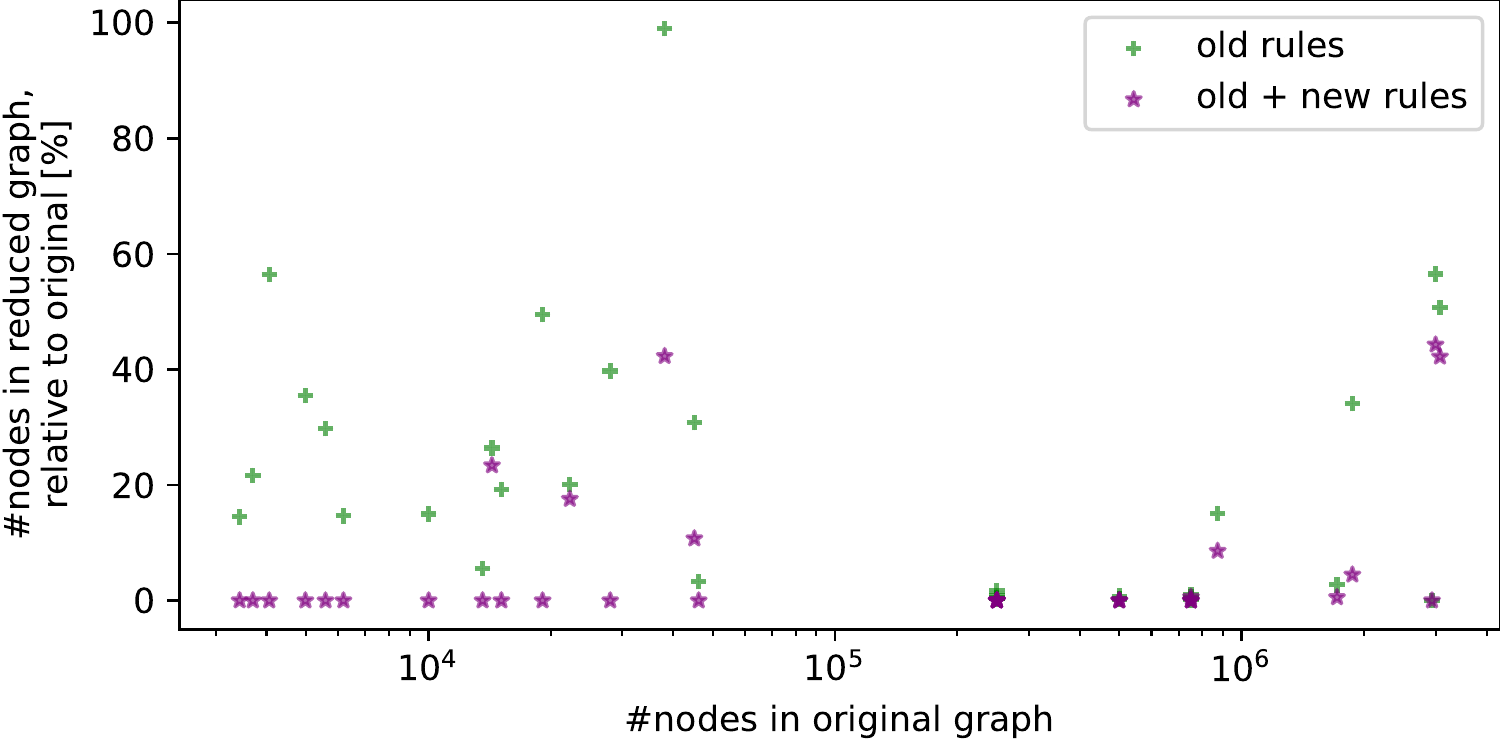}
\caption{Original graph sizes and reduced graph sizes for old and old + new reductions.}\label{fig:kernel-sizes}
\end{figure}
The largest reduction in the \texttt{REP} instance set is observed on \texttt{web-wikipedia\_link\_it}, where the new reduction rules result in an empty reduced graph, \ie, the instance is fully solved by the reductions only.
The biggest improvement occurred on \texttt{sc-TSOPF-RS-b2383}, where the old rules were barely effective and reduced the number of nodes by only roughly \SI{1}{\percent}.
In combination with the new rules, however, the computed reduced graph contains only \SI{42.29}{\percent} of the nodes of the original instance.
On all instances, using the new rules in addition to the old ones always resulted in smaller reduced graphs than when just using the old ones.
On average, the old rules alone reduced the graph size by about \SI{67}{\percent}, which improved to over \SI{80}{\percent} when combined with our new rules.

The new rules also work very well on the \texttt{RHG} instances and consistently produced smaller reduced graphs than when just using the old ones.
Generally, the reductions are very efficient on these instances.
If using only the old rules, the resulting reduced graphs are reduced to between \SI{0.14}{\percent} and \SI{1.64}{\percent} of the original graph sizes.
Combined with the new rules, the range is between \SI{0}{\percent} and \SI{0.59}{\percent}.
Two \texttt{RHG} instances were reduced to zero nodes when using the new rules in addition to the old ones.
On average, the reduced graphs obtained by old and new rules together were only \SI{0.05}{\percent} of the original graph sizes, whereas the average for the old rules alone was more than ten times larger.

The new and old rules together computed empty reduced graphs on all \texttt{OSM} instances,
which never happened when using only the rules from the literature.
On average, the old rules reduced the number of vertices down to \SI{25.4}{\percent},
where the range is relatively large and between \SI{5.58}{\percent} on \texttt{district-of-columbia-AM2} and \SI{56.42}{\percent} on \texttt{idaho-AM3}.

\emph{In summary}, our new reduction rules distinctly and consistently produce smaller reduced graphs on all \texttt{REP}, \texttt{RHG}, and \texttt{OSM} instances and even compute empty reduced graphs on \numprint{15} instances, which the old ones alone never accomplished on any instance of our collection.
Figure~\ref{fig:kernel-sizes} summarizes this visually.

\subsection{Exact Algorithms.}
\begin{table*}[t!]\centering
	\begin{tabular}{lrrrrrr}
		\toprule
		& \multicolumn{2}{c}{$t_{sol}$}& \multicolumn{2}{c}{$t_{prv}$}& \multicolumn{2}{c}{$\Weight(\hat{C})$}\\
		\cmidrule(lr){2-3}\cmidrule(lr){4-5}\cmidrule(lr){6-7}
		Instance Set&\texttt{TSM-MWC} & \texttt{MWCRedu}&\texttt{TSM-MWC} & \texttt{MWCRedu}&\texttt{TSM-MWC} & \texttt{MWCRedu} \\
        \midrule
                & \multicolumn{6}{c}{Exact Results}\\
		\midrule

		\texttt{DIMACS} & \nprounddigits{2}$\numprint{1106.98989016234}$ & \npboldmath\nprounddigits{2}$\numprint{946.338796277187}$ & \nprounddigits{2}$\numprint{1714.98112214468}$ & \npboldmath\nprounddigits{2}$\numprint{1650.88360844772}$ & \nprounddigits{0}$\numprint{6459.56906159124}$ & \npboldmath\nprounddigits{0}$\numprint{6586.57469114631}$ \\ 
		\texttt{REP} & \npboldmath\nprounddigits{2}$\numprint{117.151960879209}$ & \nprounddigits{2}$\numprint{134.603752589215}$ & \npboldmath\nprounddigits{2}$\numprint{190.448034962849}$ & \nprounddigits{2}$\numprint{259.144957174842}$ & \nprounddigits{0}$\numprint{14092.4690329748}$ & \npboldmath\nprounddigits{0}$\numprint{14320.6878536846}$ \\ 
		\texttt{RHG} & \nprounddigits{2}$\numprint{95.9268219899964}$ & \npboldmath\nprounddigits{2}$\numprint{14.5486429410803}$ & \nprounddigits{2}$\numprint{128.162743477967}$ & \npboldmath\nprounddigits{2}$\numprint{17.7322801937836}$ & \nprounddigits{0}$\numprint{106209.843553471}$ & \npboldmath\nprounddigits{0}$\numprint{106781.400506085}$ \\ 
		\texttt{OSM} & \nprounddigits{2}$\numprint{27.6198313857714}$ & \npboldmath\nprounddigits{2}$\numprint{1.54583580095575}$ & \nprounddigits{2}$\numprint{31.0145124530937}$ & \npboldmath\nprounddigits{2}$\numprint{2.43358303873101}$ & \nprounddigits{0}$\numprint{537149.366270926}$ & \npboldmath\nprounddigits{0}$\numprint{542993.419634495}$ \\ 
		\midrule
		 &\nprounddigits{2}$\numprint{136.148723585377}$ & \npboldmath\nprounddigits{2}$\numprint{41.1407699032768}$ & \nprounddigits{2}$\numprint{189.81942278599}$ & \npboldmath\nprounddigits{2}$\numprint{65.5492147085151}$ & \nprounddigits{0}$\numprint{47737.8238862274}$ & \npboldmath\nprounddigits{0}$\numprint{48359.123381942}$ \\
		 \bottomrule

		\toprule
		& \multicolumn{3}{c}{$t_{sol}$}& \multicolumn{3}{c}{$\Weight(\hat{C})$}\\
		\cmidrule(lr){2-4}\cmidrule(lr){5-7}
		Instance Set&\texttt{FastWCLq}&\texttt{SCCWalk4l}&\texttt{MWCPeel}&\texttt{FastWCLq}&\texttt{SCCWalk4l}&\texttt{MWCPeel}\\
		\midrule
                & \multicolumn{6}{c}{Heuristic Results}\\
		\midrule
		\texttt{DIMACS} & \nprounddigits{2}$\numprint{193.114083760604}$ & \npboldmath\nprounddigits{2}$\numprint{4.462994464033}$ & \nprounddigits{2}$\numprint{91.3145981302544}$ & \nprounddigits{0}$\numprint{6791.94143574621}$ & \npboldmath\nprounddigits{0}$\numprint{6967.65184784029}$ & \nprounddigits{0}$\numprint{6547.1919320318}$ \\		
		\texttt{REP} &   \nprounddigits{2}$\numprint{190.358991737439}$ & \nprounddigits{2}$\numprint{345.882674167129}$ & \npboldmath\nprounddigits{2}$\numprint{52.1176742711867}$ & \npboldmath\nprounddigits{0}$\numprint{14189.619550757}$ & \nprounddigits{0}$\numprint{9382.02148409768}$ & \nprounddigits{0}$\numprint{14055.8247224441}$ \\ 
		\texttt{RHG} &  
		\nprounddigits{2}$\numprint{40.3595498783001}$ & \nprounddigits{2}$\numprint{513.373605546655}$ & \npboldmath\nprounddigits{2}$\numprint{10.6914413797986}$ & \npboldmath\nprounddigits{0}$\numprint{106781.400506085}$ & \nprounddigits{0}$\numprint{63647.3112386311}$ & \nprounddigits{0}$\numprint{106730.806811308}$ \\
		\texttt{OSM} & 
		\nprounddigits{2}$\numprint{4.44816405727783}$ & \nprounddigits{2}$\numprint{64.2905786783677}$ & \npboldmath\nprounddigits{2}$\numprint{1.45396003537343}$ & \npboldmath\nprounddigits{0}$\numprint{542993.419634495}$ & \nprounddigits{0}$\numprint{528956.269376379}$ & \npboldmath\nprounddigits{0}$\numprint{542993.419634495}$ \\
		\midrule
		&
		\nprounddigits{2}$\numprint{50.6849022861707}$ & \nprounddigits{2}$\numprint{84.4858556416954}$ & \npboldmath\nprounddigits{2}$\numprint{16.4921961698774}$ & \npboldmath\nprounddigits{0}$\numprint{48619.8613205878}$ & \nprounddigits{0}$\numprint{38516.3987316728}$ & \nprounddigits{0}$\numprint{48056.1462984885}$ \\
		\bottomrule
		\end{tabular}
	\caption{Overview of results for exact (top) and heuristic (bottom) algorithms as geometric mean per graph set.}\label{tab:meanheur} \label{tab:meanex}
\end{table*}
We discuss the aggregated results for each of the four instance sets (see Table~\ref{tab:meanex}).

Our algorithm \texttt{MWCRedu} is more than an order of magnitude faster in the geometric mean than \texttt{TSM-MWC} on the \texttt{OSM} instances (Table~\ref{tab:osmex}), both with respect to time to find the solution $t_{sol}$ and to prove optimality $t_{prv}$.
It is also consistently faster than \texttt{TSM-MWC} on each of the twelve instances in the set.
As both are exact algorithms, the solution weights are identical except for two cases, where \texttt{TSM-MWC} failed to find the optimal solution within the time limit
and stopped prematurely with a worse result.
Thus, \texttt{MWCRedu} dominates here.

On \texttt{DIMACS} (Table~\ref{tab:dimacsex}), no major difference in performance between the two solvers is observable.
\texttt{MWCRedu} was able to finish on nine of the \num{23} instances within the time limit, whereas \texttt{TSM-MWC} finished on only eight instances.
The running times generally lie very close together, and the solution weights are identical except for seven cases.
The reason for the similar behavior is that none of the new exact reductions employed by \texttt{MWCRedu} is able to remove vertices or edges for any instance in this set.
Thus, the solver quickly proceeds to apply the \BnB{} solver, which uses the same techniques as \texttt{TSM-MWC}.
The overhead from applying the reduction rules is only notable for the easier instances.
On average over those instances, where both finished regularly, \texttt{MWCRedu} performs slightly better, which is likely due to better initial solutions obtained from running local search~during~the~reduction~phase.

On the \texttt{REP} instances (Table~\ref{tab:repex}), the results are mixed.
\texttt{MWCRedu} and \texttt{TSM-MWC} both outperform the respective other algorithm for some instances.
On three instances, \texttt{TSM-MWC} failed to prove optimality of a solution and terminated with a suboptimal result twice.
\texttt{TSM-MWC} is very efficient for large instances with more than \num{1000000} vertices, whereas \texttt{MWCRedu} outperforms \texttt{TSM-MWC} on the smaller, more dense biology graphs.

On \texttt{RHG} (Table~\ref{tab:rhgex}), \texttt{MWCRedu} outperforms its competitor \texttt{TSM-MWC} clearly.
While \texttt{TSM-MWC} runs into a timeout twice and terminates with a suboptimal solution, \texttt{MWCRedu} always finishes regularly and is the faster algorithm except on one instance.
Its dominance in running time is pronounced and up to two orders of magnitude.
The reason for \texttt{MWCRedu}'s good performance is likely the structure of the instances, which allows it to remove most vertices quickly using very efficient reductions.

\emph{In summary}, \texttt{MWCRedu} is clearly the better algorithm on the \texttt{OSM} and \texttt{RHG} instances and on par with \texttt{TSM-MWC} on the \texttt{DIMACS} graphs.
On instances that are small and dense, such as in the \texttt{REP} set, \texttt{TSM-MWC} may be the faster algorithm, whereas \texttt{MWCRedu} can play out its strengths on very large ones. %
Notably, \texttt{MWCRedu} finished within the time limit on the same instances as \texttt{TSM-MWC} plus some more, making it the \emph{more~reliable~candidate}.

\subsection{Heuristic Algorithms.}
We now compare our heuristic solver \texttt{MWCPeel} against the state-of-the-art solvers \texttt{FastWCLq} and \texttt{SCCWalk4l} and discuss the differences on each of the four instance sets. Aggregated results are presented in Table~\ref{tab:meanex}.

As shown in Table~\ref{tab:osmheur}, \texttt{MWCPeel} performs best for \num{11} out of \num{12} \texttt{OSM} instances.
Both \texttt{MWCPeel} and \texttt{FastWCLq} find the optimal solution to all instances. %

For the \texttt{DIMACS} graphs (Table~\ref{tab:dimacsheur}), \texttt{SCCWalk4l} clearly dominates its competitors.
Between \texttt{FastWCLq} and \texttt{MWCPeel}, \texttt{FastWCLq} mostly computes slightly higher weight solutions, though it takes longer to compute them.
Looking at the instances where \texttt{TSM-MWC} fails to find the optimal solution, both \texttt{FastWCLq} and \texttt{MWCPeel} achieve higher weight solutions in a much smaller amount of time for most of them.

As shown in Table~\ref{tab:repheur}, performance on \texttt{REP} graphs is very competitive among the heuristic solvers.
While all algorithms compute the best solution an approximately equal amount of times, the solution quality of \texttt{SCCWalk4l} is the lowest on average.
Taking speed into account, \texttt{MWCPeel} shows a good performance in comparison. On average, \texttt{MWCPeel} is a factor 3.7 faster than the second fastest algorithm $\texttt{FastWCLq}$ which computing 0.9\% better solutions on average than \texttt{MWCPeel}.
It should be noted, however, that our exact solver \texttt{MWCPeel} computes even higher weight solutions than $\texttt{FastWCLq}$, while also being faster on average. 

The results for \texttt{RHG} are presented in Table~\ref{tab:rhgheur}.
Here, \texttt{MWCPeel} outperforms the other solvers in \num{31} out of \num{39} instances.
While \texttt{FastWCLq} sometimes finds a slightly higher weight solution than \texttt{MWCPeel}, it has a higher running time on average (a factor 3.8).
\texttt{SCCWalk4l} is clearly outperformed both in speed and solution quality.

\section{Conclusion} 
We presented an exact algorithm called \texttt{MWCRedu} and a heuristic algorithm called \texttt{MWCPeel} for solving the maximum weight clique problem.  Our algorithms interleave successful techniques from related work with novel data reduction rules that use local graph structures to identify and remove vertices and edges while maintaining the optimal solution. In experiments on a large range of graphs, we find that they outperform the current state-of-the-art solvers on most inputs. In particular, \texttt{MWCRedu} is faster by orders of magnitude on naturally weighted, medium-sized street network graphs and random hyperbolic graphs. \texttt{MWCPeel} outperforms its competitors on these instances, but is slightly less effective on extremely dense or large instances. In future work, we want to consider parallelization of our approaches. Given the good results of our algorithms, we plan to release them as open source.

   \ifDoubleBlind
\else 
\noindent \textbf{Acknowledgments.}
We acknowledge support by DFG grant SCHU 2567/3-1.
N.\,K.~was supported by the Vienna Science and Technology Fund (WWTF) through project VRG19-009.
\fi

\bibliographystyle{plain}
\bibliography{refThesis.bib, phdthesiscs.bib}
\appendix
\onecolumn
\section{Reduced Graph Sizes}
\begin{table*}[h!]
	\centering
	\renewcommand{\arraystretch}{0.9}
\begin{tabular}{l@{\hskip 43pt}rrrr}
		\toprule
		& \multicolumn{4}{c}{Reduced Graph Size}  \\
    \cmidrule(lr){2-5}
		& \multicolumn{2}{c}{old+new reductions} & \multicolumn{2}{c}{old reductions only} \\
    \cmidrule(lr){2-3}\cmidrule(lr){4-5}
		Graph & absolute & \% of $n_0$ & absolute & \% of $n_0$ \\

		\midrule
		\multicolumn{5}{c}{\texttt{REP}}\\
		\midrule
						
		\texttt{aff-digg} & \nprounddigits{0}$\textbf{\numprint{74864.2}}$ & \nprounddigits{2}$\textbf{\numprint{8.57922445228289}}$ & \nprounddigits{0}$\numprint{131623.666666667}$ & \nprounddigits{2}$\numprint{15.0836979432866}$ \\ 
		\texttt{bio-human-gene1} & \nprounddigits{0}$\textbf{\numprint{3915.46666666667}}$ & \nprounddigits{2}$\textbf{\numprint{17.5715418330865}}$ & \nprounddigits{0}$\numprint{4485.33333333333}$ & \nprounddigits{2}$\numprint{20.128947329055}$ \\ 
		\texttt{bio-human-gene2} & \nprounddigits{0}$\textbf{\numprint{3352.66666666667}}$ & \nprounddigits{2}$\textbf{\numprint{23.3798233379823}}$ & \nprounddigits{0}$\numprint{3786.66666666667}$ & \nprounddigits{2}$\numprint{26.4063226406323}$ \\ 
		\texttt{bio-mouse-gene} & \nprounddigits{0}$\textbf{\numprint{4840}}$ & \nprounddigits{2}$\textbf{\numprint{10.7314693687501}}$ & \nprounddigits{0}$\numprint{13887.6666666667}$ & \nprounddigits{2}$\numprint{30.7923697183359}$ \\
\bottomrule
	\end{tabular}
        \caption{Reduced graph sizes (number of nodes) with when both old and new data reductions rules are applied vs. reduced graph sizes obtained when only running reductions from the current literature.
         $n_0$ refers to the initial number of nodes. Smaller is better.
				 On the \texttt{DIMACS} instances reduced graph sizes did not change.}\label{tab:kernelsizes}
\end{table*}

\clearpage
\section{Detailed Results for Exact Algorithms}
\begin{table*}[h!]\centering\begin{tabular}{lrrrrrr}

	\toprule
	& \multicolumn{2}{c}{$t_{sol}$}& \multicolumn{2}{c}{$t_{prv}$}& \multicolumn{2}{c}{$\Weight(\hat{C})$}\\
	\cmidrule(lr){2-3}\cmidrule(lr){4-5}\cmidrule(lr){6-7}
	Graph&\texttt{TSM-MWC} & \texttt{MWCRedu}&\texttt{TSM-MWC} & \texttt{MWCRedu}&\texttt{TSM-MWC} & \texttt{MWCRedu} \\
	\midrule		
        & \multicolumn{6}{c}{\texttt{OSM} Exact Results}\\
		\midrule

	\texttt{district-of-col-AM2} & \nprounddigits{2}$\numprint{0.795482}$ & \npboldmath\nprounddigits{2}$\numprint{0.153333333333333}$ & \nprounddigits{2}$\numprint{0.883069}$ & \npboldmath\nprounddigits{2}$\numprint{0.206666666666667}$ & \npboldmath\nprounddigits{0}$\numprint{235777}$ & \npboldmath\nprounddigits{0}$\numprint{235777}$ \\ 
	\texttt{district-of-col-AM3} & \nprounddigits{2}$\numprint{1937.83}$ & \npboldmath\nprounddigits{2}$\numprint{5.64}$ & \nprounddigits{2}$\numprint{1938.06}$ & \npboldmath\nprounddigits{2}$\numprint{6.51666666666667}$ & \npboldmath\nprounddigits{0}$\numprint{545969}$ & \npboldmath\nprounddigits{0}$\numprint{545969}$ \\ 
	\texttt{greenland-AM3} & \nprounddigits{2}$\numprint{10.3079}$ & \npboldmath\nprounddigits{2}$\numprint{0.82}$ & \nprounddigits{2}$\numprint{11.9071}$ & \npboldmath\nprounddigits{2}$\numprint{3.29666666666667}$ & \npboldmath\nprounddigits{0}$\numprint{604575}$ & \npboldmath\nprounddigits{0}$\numprint{604575}$ \\ 
	\texttt{hawaii-AM3} & \nprounddigits{2}$\numprint{3598.87}$ & \npboldmath\nprounddigits{2}$\numprint{30.8833333333333}$ & - & \npboldmath\nprounddigits{2}$\numprint{54.6833333333333}$ & \nprounddigits{0}$\numprint{1110978}$ & \npboldmath\nprounddigits{0}$\numprint{1229741}$ \\ 
	\texttt{idaho-AM3} & \nprounddigits{2}$\numprint{218.66}$ & \npboldmath\nprounddigits{2}$\numprint{4.46666666666667}$ & \nprounddigits{2}$\numprint{220.419}$ & \npboldmath\nprounddigits{2}$\numprint{5.55333333333333}$ & \npboldmath\nprounddigits{0}$\numprint{1101721}$ & \npboldmath\nprounddigits{0}$\numprint{1101721}$ \\
\bottomrule
                        \end{tabular}
	\caption{\texttt{OSM} and \texttt{DIMACS} exact results  for each graph and the geometric mean.}\label{tab:dimacsex}\label{tab:osmex}
	\end{table*}

\begin{table*}\centering\begin{tabular}{lrrrrrr}
	\toprule
	& \multicolumn{2}{c}{$t_{sol}$}& \multicolumn{2}{c}{$t_{prv}$}& \multicolumn{2}{c}{$\Weight(\hat{C})$}\\
	\cmidrule(lr){2-3}\cmidrule(lr){4-5}\cmidrule(lr){6-7}
	Graph&\texttt{TSM-MWC} & \texttt{MWCRedu}&\texttt{TSM-MWC} & \texttt{MWCRedu}&\texttt{TSM-MWC} & \texttt{MWCRedu} \\
	\midrule	

                & \multicolumn{6}{c}{\texttt{REP} Exact Results}\\
		\midrule
	\texttt{aff-digg} & \npboldmath\nprounddigits{2}$\numprint{16.5473833333333}$ & \nprounddigits{2}$\numprint{45.3955555555556}$ & \npboldmath\nprounddigits{2}$\numprint{244.038333333333}$ & \nprounddigits{2}$\numprint{273.673333333333}$ & \npboldmath\nprounddigits{0}$\numprint{3828.66666666667}$ & \npboldmath\nprounddigits{0}$\numprint{3828.66666666667}$ \\ 
	\texttt{bio-human-gene1} & \nprounddigits{2}$\numprint{3598.98333333333}$ & \npboldmath\nprounddigits{2}$\numprint{2010.81777777778}$ & - & \npboldmath\nprounddigits{2}$\numprint{3327.64888888889}$ & \nprounddigits{0}$\numprint{136325.333333333}$ & \npboldmath\nprounddigits{0}$\numprint{136692.111111111}$ \\ 
	\texttt{bio-human-gene2} & \nprounddigits{2}$\numprint{2298.36}$ & \npboldmath\nprounddigits{2}$\numprint{474.49}$ &- & \npboldmath\nprounddigits{2}$\numprint{1380.66111111111}$ & \npboldmath\nprounddigits{0}$\numprint{131903.666666667}$ & \npboldmath\nprounddigits{0}$\numprint{131903.666666667}$ \\ 
	\texttt{bio-mouse-gene} & \nprounddigits{2}$\numprint{1789.699}$ & \npboldmath\nprounddigits{2}$\numprint{198.465555555556}$ &- & \npboldmath\nprounddigits{2}$\numprint{240.965555555556}$ & \nprounddigits{0}$\numprint{50785.3333333333}$ & \npboldmath\nprounddigits{0}$\numprint{59475.6666666667}$ \\
\bottomrule
        \end{tabular}
	\caption{\texttt{REP} and \texttt{RHG} exact results for each graph and the geometric mean.}\label{tab:rhgex} \label{tab:repex}
\end{table*}

\clearpage
\section{Detailed Results for Heuristic Algorithms}
\begin{table*}[h!]\centering\begin{tabular}{lrrrrrr}
		\toprule
		& \multicolumn{3}{c}{$t_{sol}$}& \multicolumn{3}{c}{$\Weight(\hat{C})$}\\
		\cmidrule(lr){2-4}\cmidrule(lr){5-7}
		Graph&\texttt{FastWCLq}&\texttt{SCCWalk4l}&\texttt{MWCPeel}&\texttt{FastWCLq}&\texttt{SCCWalk4l}&\texttt{MWCPeel}\\
		\midrule

                & \multicolumn{6}{c}{\texttt{OSM} Heuristic Results}\\
		\midrule
	\texttt{district-of-columbia-AM2} & \nprounddigits{2}$\numprint{0.32}$ & \nprounddigits{2}$\numprint{4.912}$ & \npboldmath\nprounddigits{2}$\numprint{0.164}$ & \npboldmath\nprounddigits{0}$\numprint{235777}$ & \nprounddigits{0}$\numprint{234219}$ & \npboldmath\nprounddigits{0}$\numprint{235777}$ \\ 
	\texttt{district-of-columbia-AM3} & \nprounddigits{2}$\numprint{16.868}$ & \nprounddigits{2}$\numprint{208.462}$ & \npboldmath\nprounddigits{2}$\numprint{5.256}$ & \npboldmath\nprounddigits{0}$\numprint{545969}$ & \npboldmath\nprounddigits{0}$\numprint{545969}$ & \npboldmath\nprounddigits{0}$\numprint{545969}$ \\ 
	\texttt{greenland-AM3} & \nprounddigits{2}$\numprint{2.956}$ & \nprounddigits{2}$\numprint{39.334}$ & \npboldmath\nprounddigits{2}$\numprint{0.85}$ & \npboldmath\nprounddigits{0}$\numprint{604575}$ & \npboldmath\nprounddigits{0}$\numprint{604575}$ & \npboldmath\nprounddigits{0}$\numprint{604575}$ \\ 
	\texttt{hawaii-AM3} & \nprounddigits{2}$\numprint{86.472}$ & \nprounddigits{2}$\numprint{727.416}$ & \npboldmath\nprounddigits{2}$\numprint{29.822}$ & \npboldmath\nprounddigits{0}$\numprint{1229741}$ & \nprounddigits{0}$\numprint{1224690}$ & \npboldmath\nprounddigits{0}$\numprint{1229741}$ \\ 
	\texttt{idaho-AM3} & \nprounddigits{2}$\numprint{15.758}$ & \nprounddigits{2}$\numprint{162.508}$ & \npboldmath\nprounddigits{2}$\numprint{4.292}$ & \npboldmath\nprounddigits{0}$\numprint{1101721}$ & \nprounddigits{0}$\numprint{1098044}$ & \npboldmath\nprounddigits{0}$\numprint{1101721}$ \\
\bottomrule
                \end{tabular}
	\caption{\texttt{OSM} and \texttt{DIMACS} heuristic results  for each graph and the geometric mean.}
	\label{tab:osmheur}

\label{tab:dimacsheur}
\end{table*}		

\begin{table*}\centering\begin{tabular}{lrrrrrr}
		\toprule
		& \multicolumn{3}{c}{$t_{sol}$}& \multicolumn{3}{c}{$\Weight(\hat{C})$}\\
		\cmidrule(lr){2-4}\cmidrule(lr){5-7}
		Graph &\texttt{FastWCLq}&\texttt{SCCWalk4l}&\texttt{MWCPeel}&\texttt{FastWCLq}&\texttt{SCCWalk4l}&\texttt{MWCPeel}\\
		\midrule
                & \multicolumn{6}{c}{\texttt{REP} Heuristic Results}\\

		\midrule
	\texttt{aff-digg} & \nprounddigits{2}$\numprint{240.008}$ & \npboldmath\nprounddigits{2}$\numprint{30.7306666666667}$ & \nprounddigits{2}$\numprint{47.6833333333333}$ & \nprounddigits{0}$\numprint{3514.4}$ & \npboldmath\nprounddigits{0}$\numprint{3828.66666666667}$ & \npboldmath\nprounddigits{0}$\numprint{3828.66666666667}$ \\ 
	\texttt{bio-human-gene1} & \nprounddigits{2}$\numprint{719.125333333333}$ & \nprounddigits{2}$\numprint{640.917333333333}$ & \npboldmath\nprounddigits{2}$\numprint{493.361333333333}$ & \nprounddigits{0}$\numprint{136580.8}$ & \nprounddigits{0}$\numprint{136647}$ & \npboldmath\nprounddigits{0}$\numprint{136713}$ \\ 
	\texttt{bio-human-gene2} & \nprounddigits{2}$\numprint{457.561333333333}$ & \nprounddigits{2}$\numprint{534.753333333333}$ & \npboldmath\nprounddigits{2}$\numprint{89.5913333333333}$ & \nprounddigits{0}$\numprint{131763.266666667}$ & \nprounddigits{0}$\numprint{131862.066666667}$ & \npboldmath\nprounddigits{0}$\numprint{131903.666666667}$ \\ 
	\texttt{bio-mouse-gene} & \nprounddigits{2}$\numprint{593.171333333333}$ & \nprounddigits{2}$\numprint{412.127333333333}$ & \npboldmath\nprounddigits{2}$\numprint{13.4306666666667}$ & \nprounddigits{0}$\numprint{59439.2666666667}$ & \npboldmath\nprounddigits{0}$\numprint{59473.4666666667}$ & \nprounddigits{0}$\numprint{59145.8666666667}$ \\
\bottomrule
                \end{tabular}
	\caption{\texttt{REP} and \texttt{RHG} heuristic results  for each graph and the geometric mean.}
	\label{tab:rhgheur}
	\label{tab:repheur}
\end{table*}

\clearpage
\section{Detailed Properties of Instances}
\label{prop:instances}
\tiny
\sisetup{
	round-mode = places,
	round-precision = 2,
	scientific-notation=true
}

\begin{table*}[h!]\centering\begin{adjustbox}{totalheight=\textheight-5\baselineskip}\begin{tabular}{llrrr}
		\toprule
		Dataset & Instance & $\abs{V}$& $\abs{E}$&$\rho$\\
		\midrule
		\texttt{DIMACS} &
		 \texttt{brock800\_1} & $\numprint{800}$ & $\numprint{207505}$ & $\num{0.649265}$ \\ 
		&\texttt{brock800\_2} & $\numprint{800}$ & $\numprint{208166}$ & $\num{0.651333}$ \\ 
		&\texttt{brock800\_3} & $\numprint{800}$ & $\numprint{207333}$ & $\num{0.648727}$ \\ 
		&\texttt{brock800\_4} & $\numprint{800}$ & $\numprint{207643}$ & $\num{0.649696}$ \\ 
		&\texttt{C1000.9} & $\numprint{1000}$ & $\numprint{450079}$ & $\num{0.901059}$ \\ 
		&\texttt{C2000.5} & $\numprint{2000}$ & $\numprint{999836}$ & $\num{0.500168}$ \\ 
		&\texttt{C2000.9} & $\numprint{2000}$ & $\numprint{1799532}$ & $\num{0.900216}$ \\
\bottomrule
	\end{tabular}\end{adjustbox}
\caption{Instance Properties}
\label{tab:properties}
\end{table*}
\FloatBarrier

\end{document}